\documentclass{PoS}

\newcommand{\Choose}[2]{{^{#1}C_{#2}}}
\def\abar{\overline{a}}

\title{Multi-meson States in Lattice QCD }

\ShortTitle{Multi-meson States in Lattice QCD }

\author{\speaker{William Detmold}%
  \thanks{On behalf of the NPLQCD collaboration. NT@UW-08-21}\\
  Department of Physics, University of Washington, Seattle, WA 98195-1560, USA\\
  E-mail: \email{wdetmold@phys.washington.edu}}

\abstract{In this contribution, I summarise the studies of the
  properties of Bose-Einstein condensed systems composed of up to
  twelve pions or kaons carried out by the NPLQCD collaboration. These
  investigations have provided precise determination the I=2 $\pi\pi$
  and I=1 $KK$ scattering lengths and the first determination of
  three-hadron interactions from QCD, finding a repulsive three-pion
  interaction of size consistent with naive dimensional analysis and a
  three kaon interaction consistent with zero. We have also determined
  the isospin (strangeness) density dependence of the isospin
  (strangeness) chemical potential, finding results in surprisingly
  good agreement with chiral perturbation theory.}

\FullConference{The XXVI International Symposium on Lattice Field Theory\\
		 July 14-19 2008\\
		 Williamsburg, Virginia, USA}

\begin{document}

\section{Many body lattice QCD}
\label{sec:many-body-lattice}

Lattice QCD has had major impact in many aspects of particle physics
phenomenology and in describing the spectra and structure of single
hadrons. Computing resources and lattice algorithms have reached a
stage where it is now worthwhile to consider more complicated hadronic
observables such as those in the baryon number, $B>1$ sector --- the
realm of nuclear physics. Here there are many observables that are
phenomenologically important to nuclear structure and interactions and
to nuclear astrophysics about which very little (or nothing) is known
experimentally or theoretically. Systems containing $n>2$ mesons are
also of interest in a number of areas from RHIC to neutron
stars. These systems present a significant opportunity for
contributions from lattice QCD. Recently, the first attempts to study
systems of more than two hadrons have been made by the NPLQCD
collaboration \cite{Beane:2007es,Detmold:2008fn,Detmold:2008yn}. The
results of these studies are summarised herein.

\section{Multi-meson systems}
\label{sec:multi-meson-systems}

It has long been known how to exploit the volume dependence of the
eigen-energies of two hadron systems to extract infinite volume
scattering phase shifts \cite{Luscher:1986pf} provided that the
effective range of the interaction, $r$ is small compared to the box
size $L$ (since $r\sim m_\pi^{-1}$ for most interactions, this
constraint is $m_\pi\ L \gg1$). In recent works, this has been
extended to systems involving $n>2$ bosons
\cite{Beane:2007qr,Tan:2007bg,Detmold:2008gh} and $n=3$ fermions
\cite{Luu} in the case when the relevant scattering length, $a$, is
also small compared to the box size.  The resulting shift in energy of
$n$ particles of mass $M$ due to their interactions is
\begin{eqnarray}
 \Delta E_n &=&
  \frac{4\pi\, \abar}{M\,L^3}\Choose{n}{2}\Bigg\{1
-\left(\frac{\abar}{\pi\,L}\right){\cal I}
+\left(\frac{\abar}{\pi\,L}\right)^2\left[{\cal I}^2+(2n-5){\cal J}\right]
\nonumber 
\\&&\hspace*{2cm}
-
\left(\frac{\abar}{\pi\,L}\right)^3\Big[{\cal I}^3 + (2 n-7)
  {\cal I}{\cal J} + \left(5 n^2-41 n+63\right){\cal K}\Big]
\nonumber
\\&&\hspace*{2cm}
+
\left(\frac{\abar}{\pi\,L}\right)^4\Big[
{\cal I}^4 - 6 {\cal I}^2 {\cal J} + (4 + n - n^2){\cal J}^2 
+ 4 (27-15 n + n^2) {\cal I} \ {\cal K}
\nonumber\\
&&\hspace*{4cm}
+(14 n^3-227 n^2+919 n-1043) {\cal L}\ 
\Big]
\Bigg\}
\nonumber\\
&&
+\ \Choose{n}{3}\left[\ 
{192 \ \abar^5\over M\pi^3 L^7} \left( {\cal T}_0\ +\ {\cal T}_1\ n \right)
\ +\ 
{6\pi \abar^3\over M^3 L^7}\ (n+3)\ {\cal I}\ 
\right]
\nonumber\\
&&
+\ \Choose{n}{3} \ {1\over L^6}\ \overline{\overline{\eta}}_3^L\ 
\ \ + \ {\cal O}\left(L^{-8}\right)
\ \ \ \ ,
\label{eq:energyshift}
\end{eqnarray}
where the parameter $\abar$ is related to the scattering length, $a$,
and the effective range, $r$, by
\begin{eqnarray}
a
& = & 
\overline{a}\ -\ {2\pi\over L^3} \overline{a}^3 r \left(\ 1 \ -\
  \left( {\overline{a}\over\pi L}\right)\ {\cal I} \right)\ \ .
\label{eq:aabar}
\end{eqnarray}
The geometric constants that enter into eq.~(\ref{eq:energyshift}) are
\begin{eqnarray*} 
  &{\cal I}\ =\ -8.9136329\,, &{\cal J}\ =\ 16.532316\,, \qquad\qquad
  {\cal K}\ = \ 8.4019240\,,
  \nonumber\\
  &{\cal L}\ = \ 6.9458079\,, &{\cal T}_0\ = -4116.2338\,,
  \qquad\qquad {\cal T}_1\ = \ 450.6392\,, 
\end{eqnarray*}
and $^nC_m=n!/m!/(n-m)!$.  The three-body contribution to the
energy-shift given in eq.~(\ref{eq:energyshift}) is represented by the
parameter $\overline{\overline{\eta}}_3^L$, which is a combination of
the volume-dependent, renormalization group invariant quantity,
$\overline{\eta}_3^L$, and contributions from the two-body scattering
length and effective range,
\begin{eqnarray}
\overline{\overline{\eta}}_3^L & = & \overline{\eta}_3^L  \left(\ 1 \ -\
  6  \left({\overline{a}\over\pi L}\right)\ {\cal I} \right)
\ +\ {72\pi \overline{a}^4 r\over M L} \ {\cal I}
\ \ \ ,
\label{eq:eta3barbar}
\end{eqnarray}
where
\begin{eqnarray*} 
  \overline{\eta}_3^L & = \eta_3(\mu)\ +\ {64\pi a^4\over
    M}\left(3\sqrt{3}-4\pi\right)\ \log\left(\mu L\right)\ -\ {96
    a^4\over\pi^2 M} {\cal S}_{\rm MS} \ \ \ .
\label{eq:etathreebar}
\end{eqnarray*}
The quantity $\eta_3(\mu)$ is the coefficient of the three-$\pi^+$
interaction that appears in the effective Hamiltonian density
describing the system \cite{Detmold:2008gh}. It is renormalization
scale, $\mu$, dependent.  The quantity ${\cal S}$ is renormalization
scheme dependent and we give its value in the minimal subtraction (MS)
scheme, ${\cal S}_{\rm MS}\ = \ -185.12506$.

Lattice QCD measurements of these energy shifts allow one to extract
the parameters $\abar$ and $\overline{\overline{\eta}}_3^L$. To
determine the energy shifts, we study the correlators (specifying to
the multi-pion system)
\begin{eqnarray}
C_n(t) 
 & \propto & \langle 
\left(\sum_{\bf x} \pi^-({\bf x},t)
\right)^n
\left( 
\phantom{\sum_x\hskip -0.2in}
\pi^+({\bf 0},0)
\right)^n
\rangle\
 \ \ ,
\label{eq:Cnfun}
\end{eqnarray}
where $\pi^+({\bf x},t)=\overline{u}({\bf x},t)\gamma_5 d({\bf x},t)$.
On a lattice of infinite temporal extent,\footnote{Effects of temporal
  (anti-)periodicity are discussed in Ref.~\cite{Detmold:2008yn}.} the
combination
\begin{eqnarray}
G_n(t) 
&  \equiv & { C_n(t) \over \left[\ C_1 (t)\ \right]^n }
\ \stackrel{t\to\infty}{\longrightarrow}\ {\cal B}_0^{(n)}\ e^{- \Delta E_n\ t}
\ \ \ ,
\label{eq:Gnlarget}
\end{eqnarray}
where $\Delta E_n$ is the energy shift appearing in
Eq.~(\ref{eq:energyshift}).  

To compute the $(n!)^2$ Wick contractions in Eq.~(\ref{eq:Cnfun}), we
note that this correlation function can be written as
\begin{eqnarray}
C_n(t) 
 & \propto & 
\langle \ \left(\ \overline{\eta} \Pi \eta\ \right)^n \ \rangle
 \ \ ,
\label{eq:Cnfungrassman}
\end{eqnarray}
where
\begin{eqnarray}
  \Pi
&=& \sum_{\bf x} \ S({\bf x},t;0,0) \   S^\dagger({\bf x},t;0,0)
\ \ \  ,
\label{eq:PiDef}
\end{eqnarray}
and $S({\bf x},t;0,0)$ is a light-quark propagator.  The object
(block) $\Pi$ is a $12\times 12$ (4-spin and 3 color) bosonic
time-dependent matrix, and $\eta_\alpha$ is a twelve component
Grassmann variable.  Using
\begin{eqnarray}
\langle \overline{\eta}^{\alpha_1} \overline{\eta}^{\alpha_2}... 
\overline{\eta}^{\alpha_n} 
\eta_{\beta_1} \eta_{\beta_2} ... \eta_{\beta_n} \rangle
& \propto & 
\varepsilon^{\alpha_1\alpha_2..\alpha_n\xi_1..\xi_{12-n}}\ 
\varepsilon_{\beta_1\beta_2..\beta _n\xi_1..\xi_{12-n}}\ 
\ \ \  ,
\label{eq:GrassCon}
\end{eqnarray}
leads to correlation functions
\begin{eqnarray}
C_n(t) 
 & = & 
\varepsilon^{\alpha_1\alpha_2..\alpha_n\xi_1..\xi_{12-n}}\ 
\varepsilon_{\beta_1\beta_2..\beta _n\xi_1..\xi_{12-n}}\ 
\left(\Pi\right)_{\alpha_1}^{\beta_1} \left(\Pi\right)_{\alpha_2}^{\beta_2} 
.. \left(\Pi\right)_{\alpha_n}^{\beta_n} 
\ \ \  .
\label{eq:Cnepep}
\end{eqnarray}
For an arbitrary $12\times 12$ matrix, $A$,
\begin{eqnarray}
\det\left(1+\lambda A\right) & = & 
{1\over 12!}\ 
\varepsilon^{\alpha_1\alpha_2..\alpha_{12}}\ 
\varepsilon_{\beta_1\beta_2..\beta_{12}}\ 
\left(1+\lambda A\right)_{\alpha_1}^{\beta_1} 
\left(1+\lambda A\right)_{\alpha_2}^{\beta_2} 
\ldots\left(1+\lambda A\right)_{\alpha_{12}}^{\beta_{12}} 
\nonumber\\
& = & 
{1\over 12! }\ \left[\ 
\varepsilon^{\alpha_1\alpha_2..\alpha_{12}}\ 
\varepsilon_{\alpha_1\alpha_2..\alpha_{12}}\ 
\ +\ 
\lambda \ \Choose{12}{1}\ 
\varepsilon^{\alpha_1\alpha_2..\alpha_{12}}\ 
\varepsilon_{\beta_1\alpha_2..\alpha_{12}}\ 
\left(\ A\ \right)_{\alpha_1}^{\beta_1}
+ \ldots
\right.
\nonumber\\
& & \left. 
\qquad\ +\ 
 \lambda^n \ \Choose{12}{n}\ 
\varepsilon^{\alpha_1\alpha_2..\alpha_n\xi_1..\xi_{12-n}}\ 
\varepsilon_{\beta_1\beta_2..\beta _n\xi_1..\xi_{12-n}}\ 
\left(\ A\ \right)_{\alpha_1}^{\beta_1} 
\left(\ A\ \right)_{\alpha_2}^{\beta_2} 
\ldots \left(\ A\ \right)_{\alpha_n}^{\beta_n} 
\right.
\nonumber\\
& & \left.\qquad
\ \ldots\ \ +\ 
\lambda^{12} 
\varepsilon^{\alpha_1\alpha_2..\alpha_{12}}\ 
\varepsilon_{\beta_1\beta_2..\beta_{12}}\ 
\left(\ A\ \right)_{\alpha_1}^{\beta_1} 
\ldots \left(\ A\ \right)_{\alpha_{12}}^{\beta_{12}} 
\ \right]
\nonumber\\
&=&
\frac{1}{12!}\ \sum_{j=1}^{12}\ \Choose{12}{j} \ \lambda^j\  C_j(t)
\ \ \ ,
\label{eq:detA}
\end{eqnarray}
where in the last line we identify the matrix $A$ with $\Pi$.
Further,
\begin{eqnarray}
\det\left(1+\lambda A\right) & = & 
\exp\left({\rm Tr}\left[  \log\left[\ 1+\lambda A\right]\ \right]\ \right)
\ =\ 
\exp\left({\rm Tr}\left[ \sum_{p=1} {(-)^{p-1}\over p} \lambda^p A^p \right]\
\right)\
\nonumber\\
& = &
1\ +\ \lambda\ {\rm Tr}\left[\ A\ \right]
\ +\ 
{\lambda^2\over 2}\ \left(\ 
\left(  {\rm Tr}\left[\ A\ \right] \right)^2
\ -\ 
{\rm Tr}\left[\ A^2\ \right] 
\right)
\nonumber\\
& &\ +\ 
{\lambda^3\over 6}\ \left(\ 
2 {\rm Tr}\left[\ A^3\ \right] \ -\ 
3  {\rm Tr}\left[\ A\ \right]   {\rm Tr}\left[\ A^2\ \right] \ +\ 
 \left(\ {\rm Tr}\left[\ A\ \right] \right)^3\ \right)
\ +\ \ldots
\ .
\label{eq:detB}
\end{eqnarray}
Therefore, by equating terms of the same order in the expansion
parameter $\lambda$ in Eq.~(\ref{eq:detA}) and Eq.~(\ref{eq:detB}),
one can recover the $n$-$\pi^+$ correlation functions in
Eq.~(\ref{eq:Cnepep}).  As an example, the contractions for the
$3$-$\pi^+$ system are
\begin{eqnarray}
C_3(t) & \propto & 
{\rm tr_{C,S}}\left[ \Pi \right]^3
\ -\  3\  {\rm tr_{C,S}}\left[ \Pi^2 \right] {\rm tr_{C,S}}\left[\Pi\right]
\ +\  2\  {\rm tr_{C,S}}\left[ \Pi^3 \right]
\ \ \ ,
\label{eq:threePiCorrelator}
\end{eqnarray}
where the traces, ${\rm tr_{C,S}}$, are over color and spin indices.
Contractions for $n\le12$ mesons are given explicitly in
Ref.~\cite{Detmold:2008fn}.

\section{Two- and three- body interactions}
\label{sec:two-three-body}

\begin{figure}[b]
  \centering
  \includegraphics[width=0.45\columnwidth]{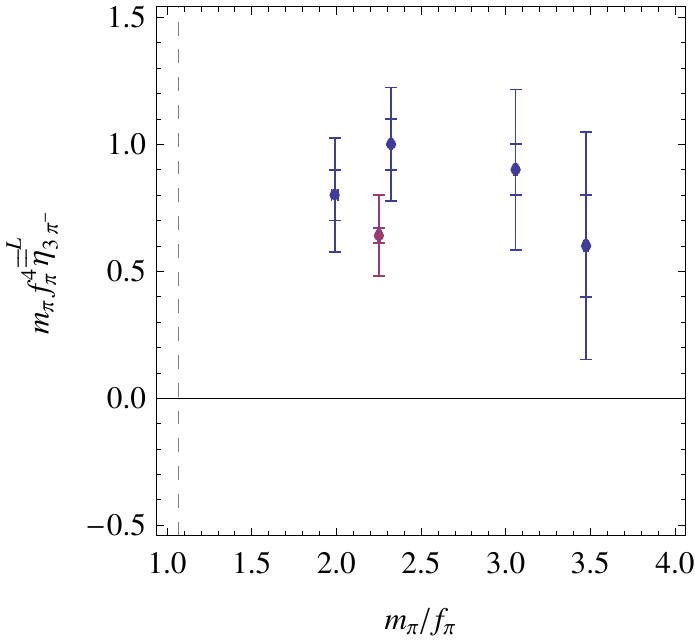}
\qquad
  \includegraphics[width=0.45\columnwidth]{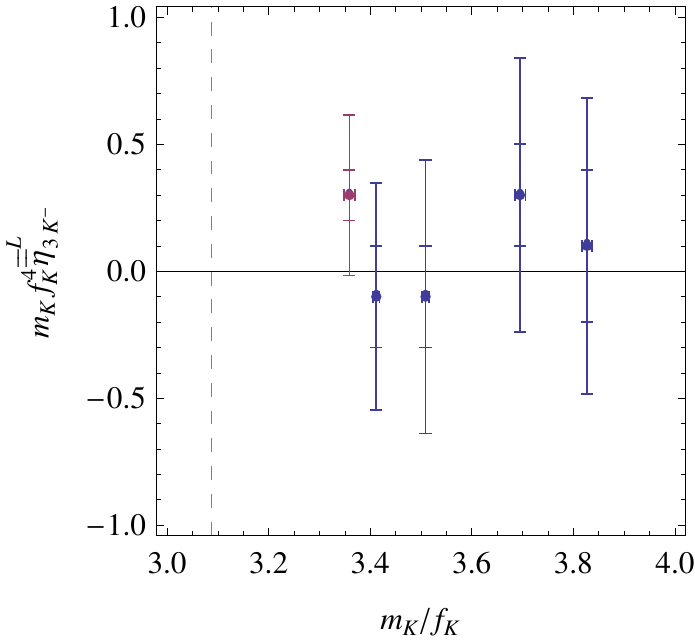}
  \caption{Three pion (left) and kaon (right) interactions determined
    from the MILC coarse (blue) and fine (magenta) lattices plotted . }
  \label{fig:three}
\end{figure}

The NPLQCD collaboration have computed the $n$ pion and kaon
correlators in the previous section using domain wall fermion
\cite{Kaplan:1992bt,Shamir:1993zy} propagators on various ensembles of
MILC 2+1 flavour rooted staggered gauge configurations
\cite{Bernard:2001av} (parameters are shown in Table
\ref{tab:MILCcnfs} and further details are given in
Refs.~\cite{Beane:2007es,Detmold:2008fn,Detmold:2008yn}). In order to
correctly calculate these correlators for large $n$, very high
numerical precision is necessary (our calculations use the {\tt
  arprec} library \cite{ARPREC}). By performing a correlated fit to
the effective energy differences extracted from these measurements, we
have determined the two- and three-body interactions. The two body
interactions extracted from this analysis agree with those extracted
from the two-body sector alone \cite{Beane:2007xs}. The resulting
three body interactions are displayed in Fig.~\ref{fig:three}. The
three pion interaction is found to be repulsive with a magnitude
consistent with the expectation from naive dimensional analysis. In
contrast, the three $K^+$ interaction is consistent with zero within
somewhat larger uncertainties.
\begin{table}[h]
  \caption{The parameters of the MILC gauge configurations and
    domain-wall quark propagators used in these calculations. The subscript $l$
    denotes light quark (up and down), and  $s$ denotes the strange
    quark. The superscript $dwf$ denotes the bare-quark mass for the
    domain-wall fermion propagator calculation. The last column is the 
    number of configurations times the number of sources per
    configuration. For the ensembles labeled with ``P$\pm$A'', propagators
    that were periodic in the temporal direction were computed in
    addition to those with anti-periodic temporal boundary conditions.}
\label{tab:MILCcnfs}
\begin{tabular}{ccccccc}
\hline
 Ensemble        
&  $b m_l$ &  $b m_s$ & $b m^{dwf}_l$ & $ b m^{dwf}_s $  & \# of propagators   \\
\hline 
2064f21b676m007m050 &  0.007 & 0.050 & 0.0081 & 0.081   & 1038\ $\times$\ 24 \\
2064f21b676m010m050 &  0.010 & 0.050 & 0.0138 & 0.081   & 768\ $\times$\ 24 \\
2064f21b679m020m050 &  0.020 & 0.050 & 0.0313 & 0.081   & 486\ $\times$\ 24 \\
2064f21b681m030m050 &  0.030 & 0.050 & 0.0478 & 0.081   & 564\ $\times$\ 20 \\
\hline
2896f2b709m0062m031 & 0.0062 & 0.031 & 0.0080 & 0.0423  & 1001\ $\times$\ 7 \\
2896f2b709m0062m031 P$\pm$A& 0.0062 & 0.031 & 0.0080 & 0.0423 & 
 1001\ $\times$\ (1+1) \\
\hline
2864f2b676m010m050 & 0.010 & 0.050 & 0.0138 & 0.081 & 
 137 \ $\times$\ 8 \\
2864f2b676m010m050 P$\pm$A & 0.010 & 0.050 & 0.0138 & 0.081 & 
 274 \ $\times$\ (2+2) \\ \hline
\end{tabular}
\end{table}

\section{Pion and kaon condensation}
\label{sec:pion-kaon-cond}

\begin{figure}[t]
  \centering
  \includegraphics[width=0.9\columnwidth]{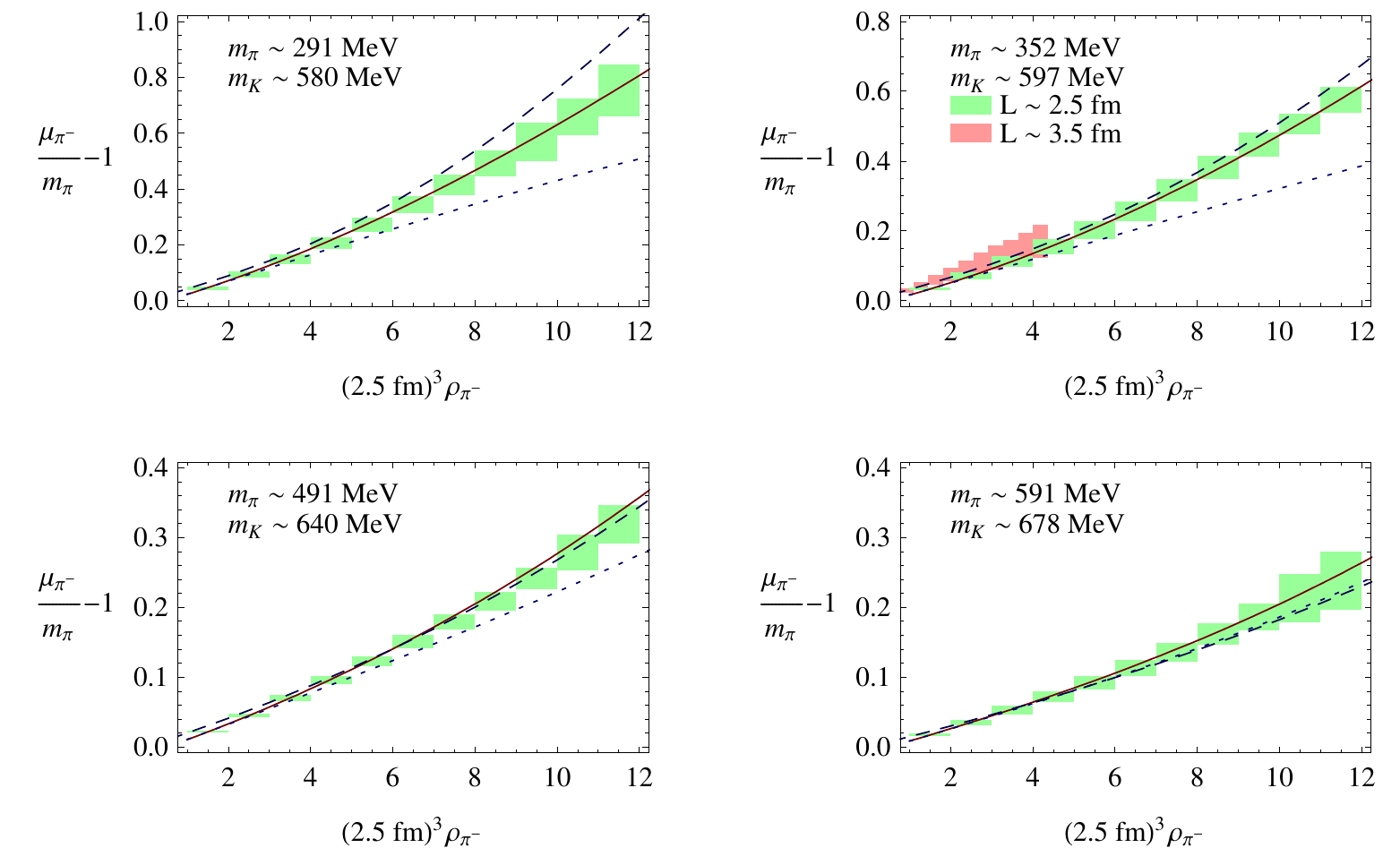}
  \caption{Dependence of the isospin chemical potential on the isospin
    density, calculated on the coarse MILC ensembles. The curves
    correspond to the predictions of tree level chiral perturbation
    theory (dashed) \cite{Son:2000xc}, the energy shift of
    Eq.~(2.1) (solid)
    and with the three-body interaction removed (dotted). }
    \label{fig:iso}
\end{figure}

\begin{figure}[h]
   \centering
   \includegraphics[width=0.9\columnwidth]{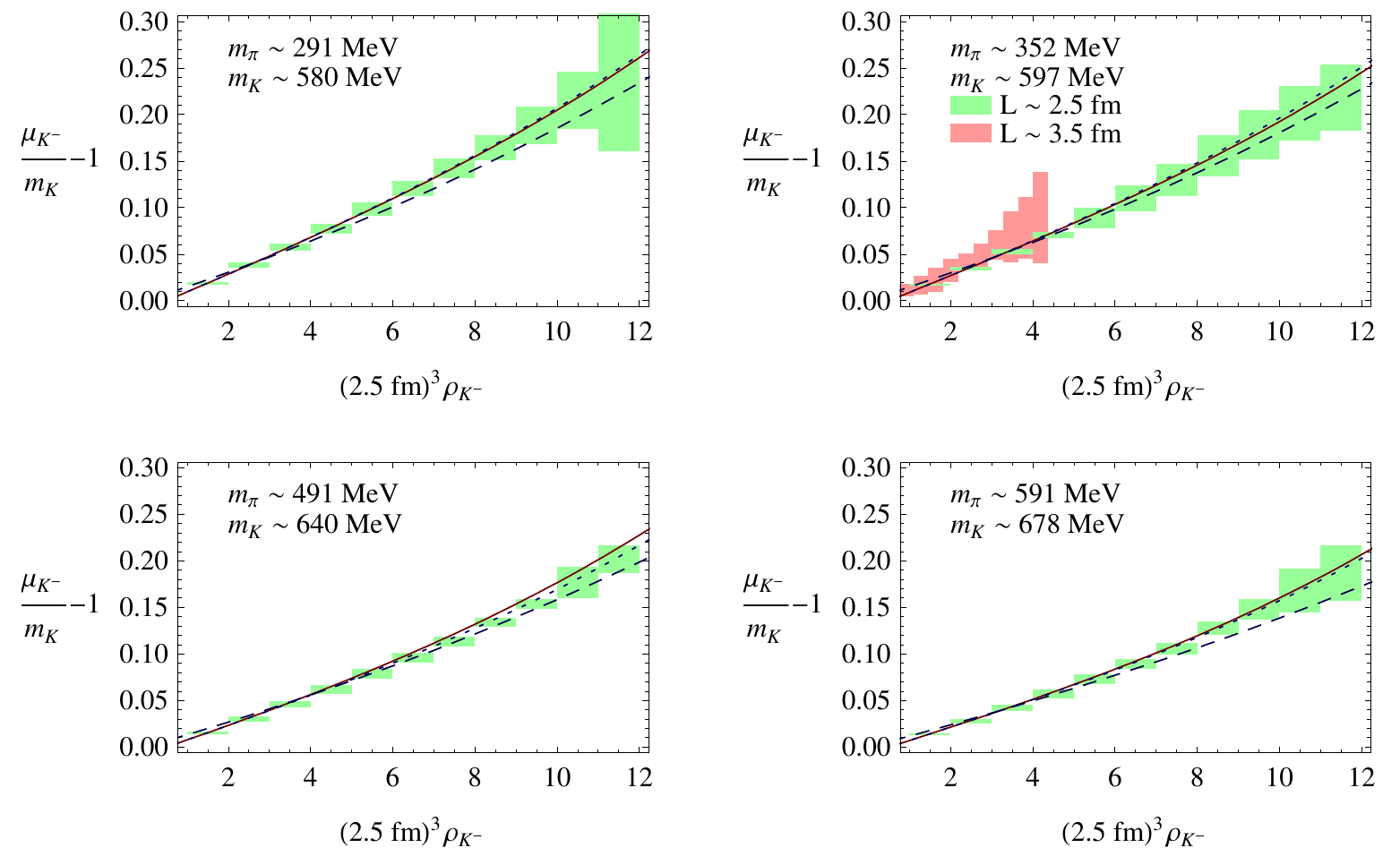}
   \caption{Dependence of the strangeness chemical potential on the
     kaon density. Details are as in Fig.~1.}
   \label{fig:str}
 \end{figure}

The ground state of the $n$ meson systems that are being studied is a
Bose-Einstein condensate of fixed $z$ component of isospin (and
strangeness in the case of kaons). It is of great interest to
investigate the properties of such systems. Theoretical efforts have
used leading order chiral perturbation theory to investigate the phase
diagram at low chemical potential \cite{Son:2000xc} and it is
important to assess the extent to which these results agree with QCD.
Our numerical calculations allow us to probe the dependence of the
energy on the pion (kaon) density, and thereby extract the chemical
potential via a finite difference. The results using the coarse MILC
lattice are shown for the pion and kaon systems in Figs.~\ref{fig:iso}
and \ref{fig:str}. Also shown is the prediction from tree-level chiral
perturbation theory, with which we find surprisingly good
agreement. This is encouraging for studies of kaon condensation in
neutron stars where, typically, tree level chiral perturbation theory
interactions are assumed amongst kaons and between kaons and baryons

\section{Summary}
\label{sec:summary}

Multi-meson systems (and in general multi-hadron systems) have been
investigated using lattice QCD. The calculations presented here
provide a first insight into the nature of these systems, but much
remains to be studied. Recently, we have started to explore the
effects of these condensed systems on other observables, looking at
how the pion condensate screens the potential between a static
quark--anti-quark pair \cite{Detmold:2008bw}.

\acknowledgments We thank R.~Edwards and B.~Joo for help with the
QDP++/Chroma programming environment~\cite{Edwards:2004sx}. This work
was supported by the the U.S. Department of Energy under Grant
No. DE-FG03-97ER4014.

\end{document}